\pdfoutput=1

\documentclass[alpha-refs]{wiley-article} %

\usepackage[english]{babel}
\usepackage{indentfirst}
\usepackage{geometry}
\usepackage{array}
\usepackage{amssymb}
\usepackage[colorlinks, citecolor=blue]{hyperref}
\hypersetup{
colorlinks=true,
linkcolor=black
}

\RequirePackage{xcolor}
 
\newcommand{\cfootnote}[2][black]{%
    {\color{#1}\footnote{#2}}%
}

\usepackage{setspace}

\usepackage{graphicx} 
\usepackage{float} 
\usepackage{subfigure} 

\papertype{Original Article}
\paperfield{Journal Section}

\title{Can Education Motivate Individual Health Demands? Dynamic Pseudo-panel Evidence from China’s Immigration}

\author[1\authfn{1}]{Shixi Kang}
\author[1\authfn{1}]{Jingwen Tan}

\contrib[\authfn{1}]{Equally contributing authors.}

\affil[1]{School of economics, Henan University, Kaifeng, Henan, 475000, China}

\corraddress{Shixi Kang, School of economics, Henan University, Kaifeng, Henan, 475000, China}
\corremail{ksx@henu.edu.cn}

\runningauthor{Shixi Kang, Jingwen Tan}

\begin{spacing}{1.5}

\end{spacing}
\begin{document}

\begin{frontmatter}
\maketitle

\begin{abstract}
\small { Enhancing residents' willingness to participate in basic health services is a key initiative to optimize the allo-cation of health care resources and promote equitable improvements in group health. This paper investigates the effect of education on resident health record completion rates using a system GMM model based on pseudo-panel that consisting of five-year cross-sectional data. To mitigate possible endogeneity, this paper controls for cohort effects while also attenuating dynamic bias in the estimation from a dynamic perspective and provides robust estimates based on multi-model regression. The results show that (1) education can give positive returns on health needs to the mobile population under the static perspective, and such returns are underestimated when cohort effects are ignored; (2) there is a significant cumulative effect of file completion rate under the dynamic perspective, and file completion in previous years will have a positive effect on the current year. (3)The positive relationship between education and willingness to make health decisions is also characterized by heterogeneity by gender, generation, and education level itself. Among them, education is more likely to promote decision-making intentions among men and younger groups, and this motivational effect is more significant among those who received basic education.}

\end{abstract}
\end{frontmatter}

\section{Introduction}\label{sec1}

\indent \setlength{\parindent}{1em}In the past decades, health economics related to health equity has been a popular research area in welfare states. In recent years, how to effectively utilize health resources, improve national health awareness, and thus promote social health benefits has also gradually become an urgent concern for developing countries. With the expansion of China's mobile population (Figure 1), the potential role of this group for economic development cannot be ignored. However, the health status of the mobile population is still worrying compared to that of the local population. As an important human resource driving economic growth, the health and equality of the migrant population needs urgent attention.

Health is universally defined as a state of mental and physical well-being and a tangible expression of social well-being, which implies that a person has a physical body capable of satisfying survival along with basic cognitive and emotional abilities \citep{hahn2015education}. Health equity is defined as the equal opportunity for every individual to improve their health status without the need and ability to access health resources being dislocated based on differences such as socioeconomic class, gender or race \citep{aday1984overview,whitehead1991concepts}. However, China has a high level of health inequality, with large disparities in health care resources between regions (Figure 2). This loss of equality can lead to disproportionate ability gaps between individuals, which in turn can bring about a cycle of relative poverty or even poverty \citep{sen2004health}.

The key to achieving this equity is to increase public spending on education. As a process of improving an individual's knowledge, values and cognitive abilities, the health-enhancing effects of education are manifested in the stimulation of an individual's need for health, which accumulates and grows over the course of his or her life \citep{mirowsky2005education}. Therefore, education spending should be seen as an important component of public sector implementation of health building and interventions, and such investments are critical to breaking the cycle of poverty for disadvantaged groups and thus reducing their health disadvantage\citep{cohen2013education}. And heterogeneity is an issue that cannot be ignored when estimating returns to education \citep{hahn2015education}. Because of the broad nature of the concept of education itself, the consequences it entails will also contain more possibilities due to sample differences. Furthermore, this feature also enhances the plausibility of causal inferences\citep{susser1973causal}.

To explore the incentive effect of education on the health needs of the mobile population and hence health equity, this study uses a pseudo-panel approach to estimate the Chinese Mobile Population Dynamic Surveillance Data (CMDS). To weaken endogeneity due to dynamic bias, we introduce a systematic GMM model to investigate the returns to health in the long run. In addition, we demonstrate the robustness of the health returns by presenting regression results from multiple models and adjustments for instrumental variables in the systematic GMM model. The findings suggest that higher levels of education can significantly increase the rate of health record completion among residents. In a dynamic-static perspective, each unit increase in education level increases the completion rate by 6.42$\%$ to 7.97$\%$, and this increase is characterised by gender and generational heterogeneity.

The remainder of the paper is organised as follows: Part 2 provides a literature review; Part 3 describes the research methodology for the pseudo-panel construction; Part 4 presents the data sources and variable selection; Part 5 reports the results of the empirical analysis; and Part 6 gives the conclusions and recommendations.

\begin{figure}[htbp]
\centering 
\begin{minipage}[b]{0.45\textwidth} 
\centering 
\includegraphics[width=1\textwidth]{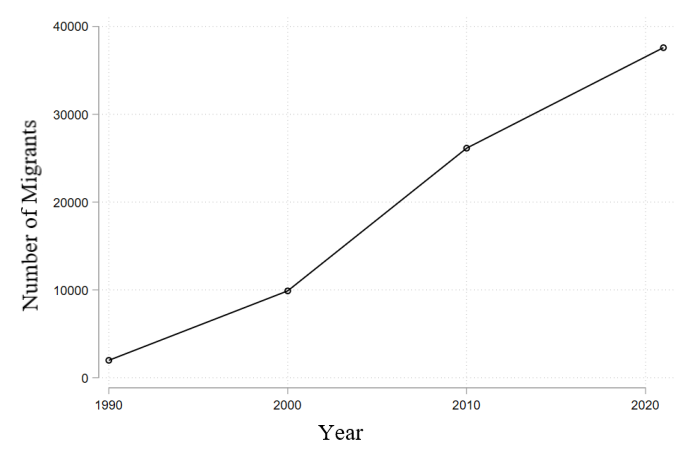}
\caption{Changes in the size of the migrants \\ \scriptsize{Note: data from National Bureau of Statistics}}
\label{Fig.1}
\end{minipage}
\begin{minipage}[b]{0.45\textwidth} 
\centering 
\includegraphics[width=1\textwidth]{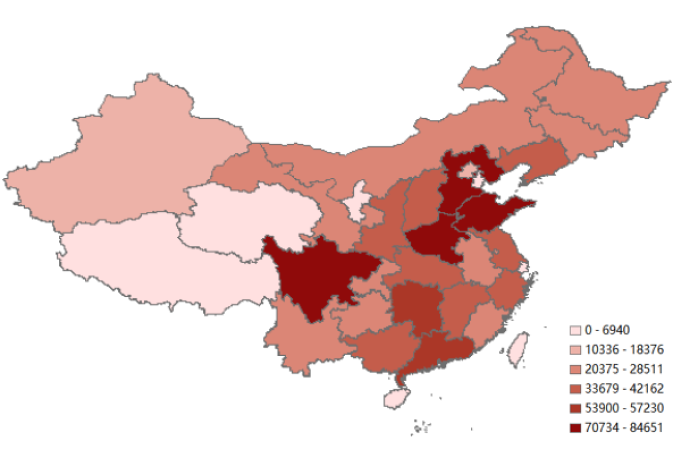}
\caption{2019 medical institution distribution \\ \scriptsize{Note: data from National Bureau of Statistics }}
\label{Fig.2}
\end{minipage}
\end{figure}

\section{Literature Review}\label{sec2}

\indent \setlength{\parindent}{1em}The positive relationship between education and health has been the subject of much scholarly research;\cite{centers2008health} found that higher grade point averages tended to be accompanied by lower rates of risky behaviours in a study of high school students in the United States. Other studies have also provided persuasive findings on the positive role of years of education in individual health risk avoidance behaviour \citep{cutler2010understanding, winkleby1992socioeconomic}. In comparing the relationship between educational attainment and self-rated health in the US and Canada, \cite{prus2011comparing} showed that those with higher education were more likely to assess their own health well, and \cite{feinstein2004contribution} found that adult education significantly increased the probability of quitting smoking and the frequency of exercise among adults to improve their health.\\
\indent \setlength{\parindent}{1em}Health estimates for specific subjects are also one of the hotspots of research. \cite{ross2010gender} showed that the mitigation of health impairment by higher levels of education was more pronounced in the female population. However, in the less educated sample, women's health was conversely weaker than men's. There is also a common view that health inequalities are common across generations and over the life course \citep{west1997health}, and \cite{ross1996education} suggest that health disparities will vary with age for people with different levels of education. According to the cumulative advantage theory, the positive impact of educational attainment on health increases with age, leading to greater heterogeneity and inequality in health among older people.\\
\indent \setlength{\parindent}{1em}In a related area of research methodology, \cite{heckman2014education} use a sequential dynamic discrete model to address endogeneity in the study of the causal effects of different years of schooling on health and the associated labour market structure.\cite{dai2021inequality}, on the other hand, controlled for individual heterogeneity by constructing pseudo-panels in their study of inequality of opportunity in China. In order to weaken the possibility of biased results in the pseudo-panel due to differences in the number of individuals included in the cohort, \cite{warunsiri2010returns} use weighted least squares (WLS) to robustly estimate the returns to education in Thailand. \cite{russell2005application} separately regress the basis of cohort segmentation in the pseudo-panel as an explanatory variable, and separately Within- and between-group estimates were used to obtain the share of two-income spouses by gender and age.\\
\indent \setlength{\parindent}{1em}The subject of this paper is to explore the incentive effects of educational attainment on the demand for basic services among cohort individuals. As the different cohorts constructed in the pseudo-panel may still have unobservable heterogeneity in groups \citep{glenn2005cohort}, we control for cohort effects as a necessary choice in the pseudo-panel analysis. Most of the current studies on the causal relationship between educational attainment and health status are based on the individual perspective, and mainly use the distribution of health resources on the supply side as a measure of health equity. This paper takes an innovative approach to explore the role of education in promoting willingness to participate in basic health activities and health decision-making from a cohort perspective on the demand side, and will construct a pseudo-panel with five-year cross-sectional data to enrich the long-term estimation of health inequalities in China, with a view to providing a new policy perspective for the optimal allocation of basic health services.

\section{Method}\label{sec3}

\indent \setlength{\parindent}{1em}Panel data play a crucial role in inferring causal relationships in the long and short term. However, an inevitable problem in the collection of panel data is the loss of tracking samples. \cite{deaton1985panel} introduced the concept of pseudo-panel data for this purpose. The cross-sectional data of a randomly selected sample at different times is used to divide the total sample into cohorts according to certain characteristics, and the mean value of each variable in each cohort is taken separately for each year of the data, and these means become the variable values of these cohorts with common characteristics, and the cohorts will replace the individuals in the original sample to form the new panel data. As pseudo-panels have the advantage of filling in the gaps in the real panel data, this paper will use a pseudo-panel treatment based on this approach for the five-year cross-sectional data used.

Assume that there is a linear function of health equality of the following form:

\begin{equation}
Health_{it} = \alpha + \beta_1edu_{it} + \beta_2X^{'} + \lambda_i + \varepsilon_{it} \quad  \quad   i = 1,...,N \quad   t = 1,...,T 
\end{equation}

${Health_{it}}$ is the degree of health equity in sample ${i}$ at time ${t}$. The paper uses the repeated question "Do you have a local health record" in the five-year CMDS data as a proxy variable for ${Health_{it}}$. $Health_{it}$ is the degree of health equity in sample $i$ at time $t$. The repeated question "Do you have a local health record" in the five-year CMDS data is used as a proxy variable for ${Health_{it}}$ and transform it into a proportional form in the pseudo-panel, i.e., the percentage of people in the group who fill out the health record. ${edu_{it}}$ is the number of years of schooling in sample ${i}$ at time ${t}$ and is the core explanatory variable in this paper. ${X{^{'}}}$contains a set of individual characteristics and urban control variables, and {${\lambda_i}$} is an unobservable individual effect.

To construct the pseudo-panel, we define a queue set $c(1,... , C)$ where each sample i will be uniquely included in the
into the cohort. These cohorts are often identified by some common characteristics of individuals, such as year of birth, race, gender, and other non-time-varying variables \citep{russell2005application}.A pseudo-panel model with a static perspective is obtained by taking the mean value of each variable in each cohort over time.

\begin{equation}
\overline{Health_{ct}} = \overline\alpha + \beta_1\overline{edu_{ct}} + \beta_2\overline{X'} + \overline{\lambda_c} + \overline{\varepsilon_{ct}} \quad  \quad   i = 1,...,N \quad   t = 1,...,T
\end{equation}

At this point, all error components associated with the original sample i are removed and the cohort effect $\overline\lambda_c$\cfootnote[black]{Since the sample size tends to differ for each year, the average effect of the cohort calculated for each year may also differ. However, the reason for not keeping a $t$ subscript for $\overline\lambda_{ct}$ is that $\overline\lambda_{ct}$ can be approximated as $\overline\lambda_{ct}$ when the number of individuals included in the cohort is sufficiently large.} substitution for the individual effect $\lambda_i$ not only controls for the effect of individual heterogeneity on the regression results, but its mean implementation form also reduces individual measurement error \citep{antman2007poverty}. Also, \cite{deaton1985panel} study showed that cohort fixed effects estimates corrected for measurement error are consistent.

However, although we attenuate endogeneity due to measurement error and omitted variables, there may still be a two-way causal relationship between education and health equity: people who are able to achieve higher levels of education inherently need good health. In addition, considering that the question "Have you established a local resident record" contains two possibilities: establishing a record in the previous year and creating a new record in the current year, in order to remove the effect of the previous year on the current year and accurately capture the causal relationship between the education level and the record completion rate of the current year's mobile population, this paper adds a first-order lag to the additional dynamic analysis of health The first-order lag term of file filling rate is added to the additional dynamic analysis in this paper to obtain a pseudo-panel model in dynamic perspective.\\
\begin{equation}
\overline{Health_{ct}} = \overline\alpha +  \beta_0\overline{Health_{ct-1}} +  \beta_1\overline{edu_{ct}} + \beta_2\overline{X'} + \overline{\lambda_c} + \overline{\varepsilon_{ct}} \quad  \quad   i = 1,...,N \quad   t = 1,...,T 
\end{equation}

A point of concern is that there is a problem that cannot be ignored when determining the number of cohorts: more cohorts increase the heterogeneity of the pseudo-panel by increasing the common characteristics of individuals, but they are also accompanied by a decrease in the number of individuals in the cohort. Therefore, identifying cohorts with a larger number of individuals is necessary for the pseudo-panel to accurately estimate subgroup means \citep{moffitt1993identification, verbeek2005estimating}. For the selection of cohorts, birth year, gender, and region were chosen as partitioning criteria in this paper in order to observe the returns of education to health equity in cohorts with generational, gender, and regional differences. Regarding the classification of birth year in the cohort, \cite{blundell1998estimating} suggested that every decade of the sample should be grouped into one cohort, while \cite{browning1985profitable} adopted the criterion of defining a household cohort every five years. Due to the large amount of data in this paper and the large number of individuals in multiple cohorts, every five years was chosen as the basis for dividing the cohorts, and a total of nine cohorts were obtained. Similarly, we divided gender (male and female) and region (east, central and west)\footnote{The geographical division in this paper is based on the concept of economic geography. Among them, the eastern region includes the eastern coastal region and the eastern three provinces, including Hebei, Beijing, Tianjin, Shandong, Jiangsu, Shanghai, Zhejiang, Fujian, Guangdong, Hainan, Jilin, Heilongjiang and Liaoning; the central region comes from the 2004 "Central Rise Plan", including Shanxi, Henan, Anhui, Jiangxi, Hunan and Hubei 6 provinces; the western region comes from the 1999 "Western Development" strategy, including Inner Mongolia, Shaanxi, Ningxia, Gansu, Xinjiang, Qinghai, Tibet, Chongqing, Sichuan, Guizhou, and Guizhou. The western region comes from the "Western Development" strategy in 1999, including 12 provinces in Inner Mongolia, Shaanxi, Ningxia, Gansu, Xinjiang, Qinghai, Tibet, Chongqing, Sichuan, Guizhou, Yunnan and Guangxi.} into 2 and 3 cohorts, respectively. The final total number of cohorts was $(9\times 2\times 3\times 4 = 216)$. In the study of \cite{verbeek2008pseudo}, the pseudo-panel estimates of subgroup means were more accurate when the number of individuals in each category was greater than 100. The demonstration of the number of individuals in each cohort in Table 1 allows us to conclude that our design is reliable.\\
\begin{table}[htb]
\renewcommand\arraystretch{1}
\begin{center}
\caption{Number of individuals in the cohort}
\begin{tabular}{m{9em} m{8em} m{8em} m{9em} m{8em}}

\hline
    Birth year & Gender & \quad & Region & \quad\\
\hline
    \quad & \quad & east & central & west\\
\hline
    1955-1959 & male & 3758 & 1142 & 3190\\
    1960-1964 & male & 7785 & 2996 & 6740\\
    1965-1969 & male & 16542 & 6388 & 13178\\
    1970-1974 & male & 25078 & 9528 & 19237\\
    1975-1979 & male & 26690 & 9969 & 18216\\
    1980-1984 & male & 33097 & 11347 & 20525\\
    1985-1989 & male & 36606 & 12821 & 23366\\
    1990-1994 & male & 21071 & 7146 & 13457\\
    1995-1999 & male & 7128 & 2241 & 5003\\
\hline
    1955-1959 & female & 2508 & 555 & 1854\\
    1960-1964 & female & 4930 & 1709 & 4015\\
    1965-1969 & female & 11366 & 4600 & 8753\\
    1970-1974 & female & 18892 & 7114 & 13366\\
    1975-1979 & female & 20842 & 7802 & 13449\\
    1980-1984 & female & 27461 & 9463 & 16328\\
    1985-1989 & female & 36750 & 13206 & 22745\\
    1990-1994 & female & 26893 & 10040 & 17875\\
    1995-1999 & female & 9626 & 3185 & 6541\\
\hline 
\end{tabular}
\end{center}
\small Note: Data are from the sum of the number of individuals in each year of CMDS 2014, 2015, 2016, 2017, and 2018.
\end{table}

\begin{table}[htb]
\renewcommand\arraystretch{1}
\begin{center}
\caption{Descriptive statistics of main variables}
\begin{tabular}{m{24em} m{3em} m{4em} m{4em} m{4em} m{4em}}

\hline
    Variables & Obs. & means & Std. & min & max\\
\hline
   Percentage of people with a health record(health) & 686,113 & 0.3546 & 0.0937 & 0.1819 & 0.6492\\
   Number of years of education(edu) & 686,113 & 10.1003 & 1.2612 & 5.8729 & 12.3210\\
   Number of people living together(living) &  686,113 & 2.9050 & 0.4553 & 1.7017 & 3.6623\\
   Mobility time(flowt) & 686,113 & 5.599776 & 2.0231 & 2.2511 & 11.0505\\
   Total household income is taken as logarithm(income) & 686,113 & 8.5697 & 0.1599 & 8.0612 & 8.9211\\
   Total number of hospitals(hos) & 686,113 & 33747.15 & 6950.85 & 24773.16 & 56054.24\\
   Total number of beds per 1,000 people(bed) & 686,113 & 5.3439 & 0.4776 & 4.5959 & 6.6144\\
   Total number of doctors per 1,000 people(doc) & 686,113 & 2.0262 & 0.2782 & 1.5312 & 2.6312\\
\hline

\end{tabular}
\end{center}
\end{table}

\section{Data}\label{sec4}
\subsection{Sources} 
\indent \setlength{\parindent}{1em}The five-year\footnote{Because of the large discrepancy in the CMDS data before and after 2014 with respect to the health status of the mobile population, only after 2014 was selected as the year of study in this paper.} repeated cross-sectional data (CMDS2014, CMDS2015, CMDS2016, CMDS2017, CMDS2018)\footnote{The original sample size of CMDS 2014 was 200937, CMDS 2015 was 206,000, CMDS 2016 was 169,000, CMDS 2017 was 169,989, and CMDS 2018 was 152,000.}of the Mobile Population Dynamic Monitoring Survey (CMDS) organized by the National Health Care Commission of China were selected as the sample for this paper. The survey covers 31 provinces (municipalities and autonomous regions) in mainland China, and uses a stratified, multi-stage, large-scale PPS sampling method to conduct a dynamic monitoring survey covering individual characteristics, employment and health status of the migrant population who have stayed in the local area for more than one month, which is an important data to measure various characteristics of the migrant population. This paper also selects some variables from the China Urban Statistical Yearbook at the provincial level\footnote{Due to the small sample of CMDS data for some prefecture-level cities in each year, if the city control variables at the prefecture-level city level are used, they cannot meet the need of estimating a large number of aggregates, and at the same time, there are more missing data for some prefecture-level cities or corps in the statistical yearbook for CMDS, so the city control variables are replaced with provincial-level data.} to control for urban characteristics.

\subsection{Variables}

\indent \setlength{\parindent}{1em}The main object of this paper is the return of education on health equity. Since the resident health record is a systematic information resource for all residents in the jurisdiction, recording their various age stages and covering various health factors in order to provide them with medical and health services, it not only measures the local capacity to provide basic health services and is an important prerequisite for residents to fully enjoy medical resources, but also represents the mobile population's own It is also an important prerequisite for residents to fully enjoy medical resources, and it can also represent the health decisions of the mobile population. Therefore, in this paper, the questionnaire "Have you established a local health record" was selected as the explanatory variable to measure the degree of health equality in each year. In this paper, "whether you have a local health record" is used as the explanatory variable to measure health equity, where the value of 1 is assigned to those who have established a record and 0 is assigned to those who have not. After pseudo-panel mean processing, this represents the percentage of people who established a file in each year, reflecting the health decision-making intentions of different groups with the same characteristics. The transformed years of education was also selected as the core explanatory variable. The transformed years of schooling was also selected as the core explanatory variable\footnote{The years of education were converted from the education level in the questionnaire. Among them, no schooling was recorded as 0; elementary school was recorded as 6; junior high school was recorded as 9; high school/junior high school was recorded as 12; university college was recorded as 15; university undergraduate was recorded as 16; and postgraduate was recorded as 19.}.

The control variables in this paper include individual strengths and urban characteristics. Due to the presence of samples with ages below 25 years, there is a possibility of updating individual education levels, so a first-order lag term for years of education is included to control for changes. Among them, the first-order lag of education level, the number of cohabitants, the duration of mobility and the logarithm of total household income are taken as individual advantage, and the total number of hospitals, the number of beds per 1,000 people and the number of practicing physicians per 1,000 people are defined as urban characteristics. Meanwhile, this paper adjusts the total household income of the sample in 2015-2018 according to the consumer price index (CPI) with 2014 as the base period, and excludes the sample with income higher than the first 2.5$\%$ and lower than the last 7.5$\%$, in order to reduce the influence of outliers on the regression results. Table 2 gives the results of descriptive statistics for the main variables after the pseudo-panel treatment. 
\section{Empirical Analysis}\label{sec5}
\subsection{Static Analysis}
\indent \setlength{\parindent}{1em}To observe the effect of cohort effects and individual differences eliminated by the pseudo-panel on the regression results, we first take a static perspective and Table 3 gives the regression results controlling for cohort effects with or without cohort effects under the pseudo-panel. Column 1 shows the OLS regression results as the reference group; column 2 gives the OLS regression results controlling for cohort effects; column 3 indicates the regression using the random effects model (REM) for the pseudo-panel; and column 4 gives the regression results for the fixed effects model (FEM) for the pseudo-panel. Since the Hausman test has shown that there is no significant correlation between the independent variables and the unobserved individual effects, we accept their original hypothesis and consider the fixed effects model as more appropriate for the main regression model in this section.\\
\indent \setlength{\parindent}{1em}With Table 3, we obtain three basic findings: (1) In the static panel, education level instantaneously gives a positive incentive to the cohort in terms of access to the same health resources: for each unit increase in education level, the rate of file completion increases by 7.97$\%$ for the mobile population group. The positive relationship between an individual's level of education and health equity has been verified in studies such as \cite{hummer2011educational} and \cite{mirowsky2017education}. This relationship is also accompanied by other characteristics, such as the tendency for such individuals to have more household cohabitation and less time on the move; (2) the negative effect of income on group file completion rates needs to be analyzed with caution. Since the survey respondents of health equity cover a wide range of ages, the effects of elderly people and children without wage income are also included in the scope of the effect of income, and the regression coefficients tend to be more underestimated at this time. (3) The regression results for Model 1 and Model 2 indicate that the inclusion of cohort fixed effects is necessary, and Model 2 has significantly higher regression coefficients on returns to education compared to Model 1, which does not include cohort effects. This finding suggests that the neglect of cohort differences can lead to serious utility underestimation. This bias is evident in the fact that the role of education is underestimated by 16.94$\%$ from both groups. One possible explanation for this phenomenon is that some people with higher levels of education tend to be engaged in complex tasks that require a lot of time, and the time occupation makes them inclined to delay their plans to fill out their health records. This negative correlation between education level and the unobservable error term could lead to an underestimation of the regression coefficient, and the introduction of cohort fixed effects in model 2 avoids this endogeneity bias. Again, this is similar to \cite{warunsiri2010returns}and \cite{juodis2018pseudo} regarding the emphasis on the importance of cohort fixed effects. Therefore, we will default to a dynamic panel analysis in the form of controlling for cohort effects.
\begin{table}[htb]
\begin{center}
\renewcommand\arraystretch{0.8}
\caption{Baseline regression results: Static analysis}
\begin{tabular}{m{11em} m{8em} m{8em} m{8em} m{8em} m{8em}}
\hline
   health & OLS(1) & OLS(2) & REM(3) & FEM(4)\\
\hline
   edu & 0.0662$^{***}$ & 0.0797$^{***}$ & 0.0675$^{***}$ & 0.0797$^{***}$ \\
   \quad  & (0.0131) & (0.0124) & (0.0128) & (0.0134) \\
   L.edu & -0.0772$^{***}$ & -0.00865 & -0.0765$^{***}$ & -0.00865 \\
   \quad  & (0.0142) & (0.0128) & (0.0119) & (0.0135) \\
   flowt & -0.00975$^{***}$ & -0.0182$^{***}$ & -0.00986$^{***}$ & -0.0182$^{***}$ \\
   \quad  & (0.00312) & (0.00486) & (0.00343) & (0.00622) \\
   income & -0.0752$^{*}$ & -0.336$^{***}$ & -0.0985$^{**}$ & -0.336$^{***}$ \\
   \quad  & (0.0427) & (0.0699) & (0.0445) & (0.0553) \\
   living & 0.117$^{***}$ & 0.200$^{***}$ & 0.130$^{***}$ & 0.200$^{***}$ \\
   \quad  & (0.0133) & (0.0157) & (0.0103) & (0.0159) \\
   hos & 4.32e-06$^{***}$ & -4.49e-06$^{***}$ & 4.06e-06$^{***}$ & -4.49e-06$^{***}$ \\
   \quad  & (0.0131) & (0.0124) & (0.0128) & (0.0134) \\
   doc & -0.221$^{***}$ & 0.0156 & -0.217$^{***}$ & 0.0156 \\
   \quad  & (0.0200) & (0.0568) & (0.0167) & (0.0634) \\
   bed & -0.0144 & -0.118$^{***}$ & -0.0227$^{***}$ & -0.118$^{***}$ \\
   \quad  & (0.00945) & (0.0194) & (0.00797) & (0.0194) \\
\hline
   Constant & 1.223$^{***}$ & 2.832$^{***}$ & 1.412$^{***}$ & 2.893$^{***}$ \\
   \quad  & (0.316) & (0.460) & (0.338) & (0.386) \\
   Cohort Fixed  &  - &  {$\surd$} & - &  {$\surd$} \\
   Cohort Number & 216 & 216 & 216 & 216 \\
   R-squared & 0.835 & 0.947 & 0.854 & 0.749 \\
\hline
\end{tabular}
\end{center}
\small {Note:t values in parentheses; $^{***}$ p<0.01, $^{**}$ p<0.05, $^{*}$ p<0.1.}
\end{table}

\begin{table}[h]
\renewcommand\arraystretch{0.8}
\begin{center}
\caption{Baseline regression results: Dynamic analysis}
\begin{tabular}{m{8em} m{7em} m{7em} m{7em} m{7em} m{7em}}
\hline
   health & OLS(1) & Diff-GMM(2) & Sys-GMM(3) & Sys-GMM(4) & Sys-GMM(5)\\
\hline
   L.health & $0.182^{***}$ & $0.211^{**}$ & $0.298^{***}$ & $0.216^{***}$ & $0.302^{**}$ \\
   \quad & (0.0581)  & (0.0909) & (0.0621) & (0.0721) & (0.128) \\
   edu & $0.0761^{***}$ & $0.0781^{***}$ & $0.0706^{***}$ & $0.0787^{***}$ & $0.0642^{***}$ \\
   \quad& (0.0121)  & (0.0177) & (0.0127) & (0.0137) & (0.0193) \\
   L.edu & -0.0209 & -0.0157 & -0.0232 & -0.0181 & -0.0289 \\
   \quad  & (0.0130) & (0.0176) & (0.0167) & (0.0167) & (0.0236) \\
   flowt & $-0.0278^{***}$ & $-0.0335^{***}$ & $-0.0441^{***}$ & $-0.0350^{***}$ & $-0.0490^{***}$\\
   \quad  & (0.00564) & (0.00978) & (0.00834) & (0.00900) & (0.0160) \\
   income & $-0.359^{***}$ & $-0.397^{***}$ & $-0.418^{***}$ & $-0.428^{***}$ & $-0.411^{***}$ \\
   \quad  & (0.0684) & (0.0878) & (0.0651) & (0.0705) & (0.115) \\
   living & $0.186^{***}$ & $0.205^{***}$ & $0.199^{***}$ & $0.206^{***}$ & $0.208^{***}$ \\
   \quad  & (0.0159) & (0.0196) & (0.0210) & (0.0214) & (0.0187) \\
   hos & $-5.88e-06^{***}$ & $-6.08e-06^{***}$ & $-6.69e-06^{***}$ & $-5.68e-06^{***}$ & $-6.38e-06^{***}$ \\
   \quad  & (1.10e-06) & (1.54e-06) & (1.32e-06) & (1.31e-06) & (1.82e-06) \\
   doc &  0.0940 & 0.124 & $0.170^{**}$ & 0.112 & 0.201 \\
   \quad  & (0.0606) & (0.0922) & (0.0749) & (0.0838) & (0.133) \\
   bed & $-0.122^{***}$ & $-0.126^{***}$ & $-0.113^{***}$ & $-0.113^{***}$ & $-0.112^{***}$ \\
   \quad  & (0.0189) & (0.0220) & (0.0193) & (0.0211) & (0.0202)\\
\hline
   Constant & $3.103^{***}$ & \quad  - & $3.537^{***}$ & $3.626^{***}$ & $3.576^{***}$ \\
   \quad  & (0.455) & \quad  - & (0.461) & (0.533) & (0.876) \\
   Cohort Number & 216 & 162 & 216 & 216 & 216\\
   Step & \quad  - & twostep & onestep & twostep & twostep \\
   IV & \quad  -  & first & second & first & second \\
   Hansen(p-value) & \quad  - & 0.229 & 0.293 & 0.913 & 0.293 \\
   AR(1)(p-value) & \quad  - & 0.000 & 0.000 & 0.000 & 0.000 \\
   AR(2)(p-value) & \quad  - & 0.967 & 0.922 & 0.913 & 0.921 \\
\hline
\end{tabular}
\end{center}
\small {Note:The first group IV is L.health, L.edu, lag(1.) collapse; edu, hos, doc, income, bed; The second group IV is L.health, L.edu, lag(1.) collapse; edu, hos, income, doc, bed, living, collapse.}
\end{table}

\subsection{Dynamic Analysis}
\indent \setlength{\parindent}{1em}Although in the static analysis we have seen a significant effect of education level on the rate of completion of health records among cohort individuals, this causal relationship is often subject to potential dynamic bias. In the question set for the dependent variable, the question form of "Have you been given a local health record" includes the possibility of being established in previous years or newly established in the current year. Therefore, this section introduces a first-order lagged term of the dependent variable in the independent variable in order to accurately capture the effect of individuals' current education level on the rate of record completion. Also, this section demonstrates the process of systematic GMM to provide robust estimates by enumerating the results of different iv under OLS, differential GMM and systematic GMM.\\
\indent \setlength{\parindent}{1em}Five different models are given in Table 4. Among them, model 1 is the reference group under OLS; models 2 and 3 are the results of two-step-based differential GMM and systematic GMM regressions under the first group of IVs, respectively; and models 3 and 5 are the results of one-step and two-step-based systematic GMM under the same IVs in the second group, respectively. Both Hansen test results, AR(1) and AR(2) values are reported in the table. As can be seen, all instrumental variables passed the overidentification test and there was no serial correlation in the model residual terms. The robustness of the results is further evidenced by the fact that the estimates of the education coefficients are convergent across the five models. The results indicate that there is long-term inertia in file completion rates and that the average level of variation is 0.2418, suggesting that the inclusion of dynamic considerations is necessary to accurately determine the returns to education in terms of health equity. At the same time, there is a significant effect of previous years' file completion rate on the current year's file completion rate.\\
\indent \setlength{\parindent}{1em}According to Table 4, the effect of the previous year sample involved in current year file completion reaches 0.182 to 0.302. After controlling for this effect, the effect of education on resident file completion rates falls back from 7.97$\%$ in the static perspective to a relatively more realistic 6.42$\%$. In addition, the estimated coefficients and significance of the other independent variables are generally consistent in the static and dynamic models, demonstrating that the estimates in the static are reliable.

\subsection{Heterogeneity Analysis}

\indent \setlength{\parindent}{1em}In addition to the overall estimate of education level on health equity returns, we still need to take into account the heterogeneous nature of such estimates. This section regresses the samples grouped according to gender, generation, and education level separately based on model 5 in Table 4. In the sample grouped by generation, we take the difference between the median year of the data, 2016, and the year of birth of the sample as the basis for determining the age of the sample, and finally divide the data into two groups using 1975, i.e., the age of the sample, as the boundary. In the sample grouped by education, we selected the years located at 70$\%$ of the overall education level, i.e., below 11 years, as the years of basic education, and above 11 years as the years of higher education.

The regression results of the heterogeneity analysis are given in Table 5. Overall, the six subgroup regressions demonstrate a significant positive effect of education level in the cohort on file completion rates. In particular, the return to education for men was 6.65$\%$, while in the female cohort, each unit increase in education level was associated with only a 4.15$\%$ increase in file completion rate. This decomposition suggests that education has a more significant return to health equity for men, similar to the findings of \cite{beckfield2018social} in their study of the gender divergence of social investment on health equity in European countries. The reason for this phenomenon may lie in the education gap between the sexes. As shown in Figure 3, in most cohorts, the female group in the sample had a lower average level of education than the male group, while \cite{ross2010gender} showed that women with low levels of education had less access to good health than men with the same level of education. In addition, subgroup regressions on generations suggest a pro-young heterogeneity characteristic of the returns to education for health equity. The return to education for health equity reached 20.2$\%$ in the sample born after 1975, while the sample before that did not show a significant benefit profile. This observation was validated in \cite{muennig2011effect} randomized controlled trial of an early education intervention with adolescents. by randomly assigning the sample to a preschool program, \cite{muennig2011effect} found that adolescents who underwent an early education intervention had better health awareness and status. Figure 4 also shows a higher level of education in the younger group compared to the older group.

We also consider the possible effect of subgroups of education level. Models 5 and 6 give the profile completion rates in each cohort under basic and higher education, respectively. For the basic education audience with less than 11 years of education, education level brings an additive profile completion rate of 5.23$\%$, while in the higher education cohort, education does not have a significant positive effect. One possible explanation is that groups with high levels of education tend to work longer hours \citep{zhang2008way}, and such forms of work apparently reduce their likelihood and willingness to fill out profiles. Although, in general, the level of education motivates people to increase their willingness to participate in health activities, this motivational effect is rather significantly reduced in the group with higher levels of education. In other words, even if high-income earners have access to more health resources, their health awareness and needs are still not promising. In contrast, even though basic education audiences are more likely to have the time to fill out health records, access to adequate and sufficient health resources is still a pressing concern. This mismatch in supply and demand between different groups constitutes a health inequity.

\begin{figure}[htb]
\centering 
\begin{minipage}[b]{0.45\textwidth} 
\centering 
\includegraphics[width=1\textwidth]{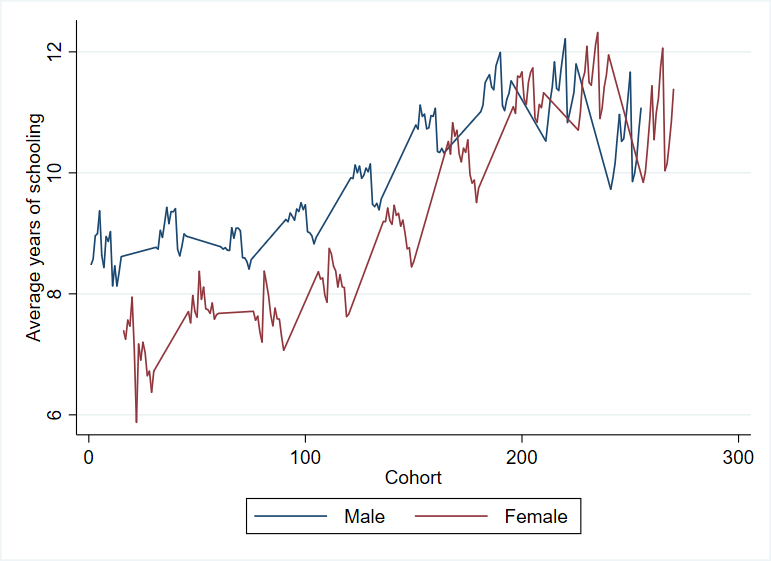} 
\caption{\\Differences in years of education from a gender perspective}
\label{Fig.3}
\end{minipage}
\begin{minipage}[b]{0.45\textwidth} 
\centering 
\includegraphics[width=1\textwidth]{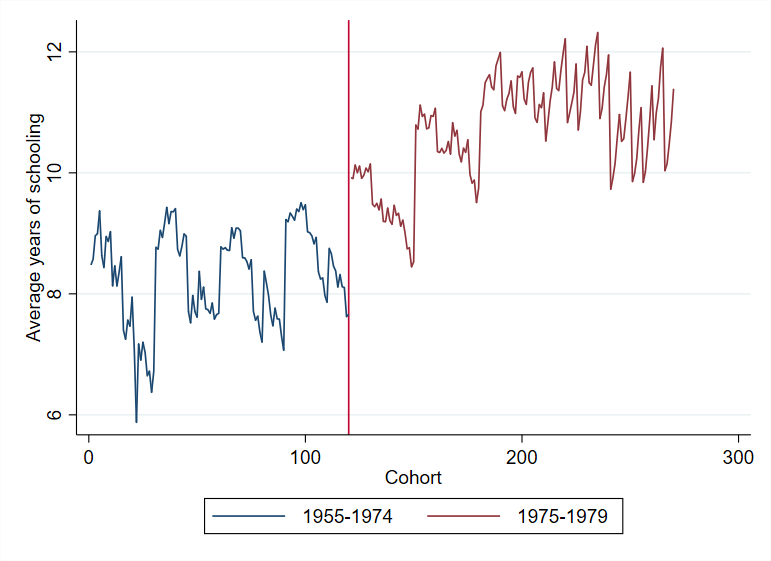}
\caption{\\ Differences in years of schooling in a generational perspective}
\label{Fig.4}
\end{minipage}
\end{figure}

\begin{table}[htb]
\renewcommand\arraystretch{0.8}
\begin{center}
\caption{Heterogeneity Analysis}
\begin{tabular}{m{7em} m{4em} m{4em} m{7em} m{7em} m{6em} m{6em}}
\hline
   health & Male(1) & Female(2) & 1955-1974(3) & 1975-1999(4) & Basic Edu(5) & Higher Edu(5) \\
\hline
   L.health & 0.168 & 0.337$^{***}$ & 0.0561 & 0.154 & 0.289$^{*}$ & 0.106 \\
   \quad & (0.158)  & (0.0973) & (0.305) & (0.170) & (0.157) & (0.392) \\
   edu & 0.0665$^{**}$ & 0.0415$^{**}$ & 0.0298 & 0.202$^{***}$ & 0.0523$^{***}$ & 0.213 \\
   \quad& (0.0316)  & (0.0190) & (0.0206) & (0.0344) & (0.0199)  & (0.167)\\
   Control Var. & \quad $\surd$ & \quad $\surd$ & \quad $\surd$ & \quad $\surd$ & \quad $\surd$ & \quad $\surd$ \\
\hline
   Constant & $3.057^{**}$ & $2.674^{**}$ & $2.546$ & $5.806^{***}$ & $2.913^{***}$ & $7.230$ \\
   \quad  & (1.520) & (1.108) & (2.545) & (0.810) & (0.841) & (4.479) \\
   Cohort Number & 108 & 108 & 72 & 72 & 162 & 54 \\
   Hansen(p-value) & 0.759 & 0.957 & 0.990 & 0.582 & 0.376 & 0.990 \\
   AR(1)(p-value) & 0.012 & 0.020 & 0.029 & 0.020 & 0.003 & 0.132 \\
   AR(2)(p-value) & 0.818 & 0.873 & 0.528 & 0.148 & 0.644 & 0.255 \\
\hline
\end{tabular}
\end{center}
\end{table}

\section{Conclusion}\label{sec6}

\indent \setlength{\parindent}{1em}This study used a systematic GMM model to estimate the health returns of group education level on the Chinese mobile population in a dynamic and static perspective based on a pseudo-panel constructed from five-year cross-sectional data of CMDS. The pseudo-panel data format weakened the estimation bias due to individual heterogeneity, and the GMM model reduced the dynamic error of education level in influencing health record completion rate. The results show that education can give transient and positive returns to health decision-making intentions to the mobile population in the static perspective, and such returns are underestimated when cohort heterogeneity is ignored; while in the dynamic perspective, there is a significant cumulative effect of file completion rate, and file completion in previous years will have a positive effect on the current year. The positive relationship between education and willingness to make healthy decisions is also characterized by heterogeneity by gender, generation, and education level itself. Among them, education was more likely to promote decision making intentions in male and younger groups, and this motivational effect was more pronounced in the group receiving basic education\\
\indent \setlength{\parindent}{1em}The high rate of return to health from education found in the study and the different returns found in the disaggregated probes urgently need to be guided by relevant policies. In the overall estimate, the increase in education levels boosted the cohort file filling rate by 6.42$\%$, and increased financial spending on education may be another measure to improve national health awareness. At the same time, for the higher level of health returns shown by the male group, while strengthening their education in order to improve overall health, attention should also be paid to the health inequalities of the female group through the gender education gap, so as to achieve true health equity in a group sense. Moreover, while we should focus on basic education, which has a greater incentive effect, this does not mean that we should invest less in higher education. Perhaps exploring the mechanisms that influence group demand for health at higher levels of education and, in turn, addressing this mechanism is a key step toward achieving health equity.


\end{document}